\documentclass[journal]{IEEEtran}

\usepackage{graphicx}
\usepackage{epsf}
\usepackage{mathtools}
\usepackage{amssymb}
\usepackage{amsmath, amsfonts}
\usepackage{latexsym}
\usepackage{multirow}
\usepackage{multicol}
\usepackage[table]{xcolor}
\usepackage{color}

\usepackage{tabularx}
\usepackage{lipsum}

\usepackage{algorithm}
\usepackage{algorithmic}

\usepackage{mathrsfs}

\usepackage{fixltx2e}

\usepackage[mathscr]{euscript}

\usepackage{commath}
\usepackage{MnSymbol}

\usepackage{subcaption}

\usepackage{mathtools} 
\DeclarePairedDelimiterX\setc[2]{[}{]}{\,#1 \;\delimsize\vert\; #2\,}
\DeclarePairedDelimiterX\parth[2]{(}{)}{\,#1 \;\delimsize\vert\; #2\,}

\PassOptionsToPackage{hyphens}{url}
\usepackage[unicode=true, bookmarks=true, bookmarksnumbered=true, bookmarksopen=true, bookmarksopenlevel=1, breaklinks=false, pdfborder={0 0 0}, pdfborderstyle={}, backref=false, colorlinks=false]{hyperref}

\usepackage{theorem}
{
\theorembodyfont{\rmfamily}

}

\usepackage{soul}

\usepackage[nospace]{cite}

\definecolor{orange}{RGB}{255,127,0}
\definecolor{blue}{RGB}{0,0,255}
\definecolor{red}{RGB}{255,0,0}
\definecolor{green}{RGB}{50,160,50}
\definecolor{grey}{RGB}{125,120,125}
\definecolor{purple}{RGB}{125,0,125}

\begin{document}
{
\title{{\fontsize{20}{2}\selectfont In-Lab Implementation of DSRC PHY Layer}}

\author
{
Leighton Thompson and Seungmo Kim, \textit{Senior Member}, \textit{IEEE}


\thanks{L. Thompson and S. Kim are with the Department of Electrical and Computer Engineering, Georgia Southern University, Statesboro, GA, USA. This work is supported by the Georgia Department of Transportation (GDOT) via grant RP21-08.}
}

\maketitle
\begin{abstract}
Connected and autonomous vehicles are already right around the corner of our everyday life. One of the key technologies actualizing the connected vehicles is vehicle-to-everything communications (V2X), which has been enhanced along the lines of two technologies--i.e., dedicated short-range communications (DSRC) and cellular V2X (C-V2X). While the United States (U.S.) federal government is on the move of transitioning to C-V2X, a vast majority of stakeholders are still heavily relying on DSRC. This makes a compelling case where DSRC still necessitates thorough studies not only theoretically but experimentally as well. To this line, we present a testbed that is dedicated to evaluating the performance of a DSRC system. Our study encompasses a few different test scenarios for precise confirmation of the performance. We claim that this experimental setup will play a vital role in future studies to assess the DSRC technology applied to various use cases.
\end{abstract}


\section{Introduction}\label{sec_intro}
\subsubsection{Background}
In the dynamic landscape of modern connectivity, the proliferation of wireless technologies has become nothing short of permeating every facet of our daily lives and redefining the way we connect, communicate, and consume information. In fact, over 55\% of website traffic comes from mobile devices, and 92.3\% of internet users access the internet using a mobile phone \cite{mobile}. In fact, a wide variety of wireless technologies immerse our daily lives, including Wi-Fi \cite{mine_lett17}, cellular technologies such as the fifth-generation (5G) cellular \cite{mine_jsac17} and the Long-Term Evolution (LTE) \cite{mine_psun}, and even wearable technologies \cite{mine_wearable_arxiv19, mine_wearable_arxiv20}.

Vehicle-to-Everything (V2X) communications \cite{mine_dsrc_arxiv20, mine_dhruba, mine_reyes} have the potential to significantly bring down the number of vehicle crashes, thereby reducing the number of associated fatalities \cite{intro1}. This capability gives V2X communications the central role in the constitution of intelligent transportation systems (ITS) for connected vehicle environments. Today, dedicated short range communications (DSRC) and cellular-V2X (C-V2X) are two key radio access technologies (RATs) that enable V2X communications. DSRC is designed to solely operate in the spectrum of 5.850-5.925 GHz (also known as the ``5.9 GHz band''), which has been earmarked in many countries for ITS applications. On the other hand, C-V2X can operate in the 5.9 GHz band as well as in the cellular operators' licensed carrier \cite{intro2}.

While the U.S. Federal Communications Commission (FCC) started endorsing C-V2X \cite{rns23_2} via waivers \cite{rns23_3} to state departments of transportation (DOTs) and other stakeholders, DSRC is still standing as the legally mandated technology \cite{rns23_1}. Moreover, DSRC has longer been earmarked in many countries for safety-critical applications. As such, the most important benefit for advocating DSRC as the key enabler of V2X communications is that it is a proven technology: it has been tested by car manufacturers for more than 10 years. Furthermore, DSRC is not bounded by patents, which requires no telecom subscription to use it. To take these advantages, since the 5.9 GHz band was dedicated for DSRC in the U.S. by the FCC in 1999, as of November 2018, more than 5,315 roadside units (RSUs) operating in DSRC were deployed nationwide \cite{intro3}. In December 2016, the National Highway Traffic Safety Administration (NHTSA) proposed to mandate DSRC for all new light vehicles \cite{intro4}.

Therefore, we suggest that DSRC still deserves in-depth investigations as a key technology realizing the connected and autonomous vehicles.

\subsubsection{Contributions}
As shall be laid out in Section \ref{sec_related}, while a rich body of literature already exists, DSRC still leaves significant room for study particularly along the line of experiment. Motivated from that, this paper lays out a preliminary study as a starting point for discussions in regard to establishing a \textit{testbed} assessing the performance of a DSRC system. As such, rather than final nor conclusive, this work should be regarded an initiative, igniting further tests and discussions on DSRC in the upcoming connected and autonomous vehicles environments. Provided that, we extend the literature on the following fronts:
\begin{itemize}
\item Presenting an \textit{in-lab testbed} that aims to evaluate the performance of digitally modulated signal exchanges
\item Providing several scenarios for testing the DSRC communications performance
\end{itemize}

\begin{figure}
\centering
\includegraphics[width = \linewidth]{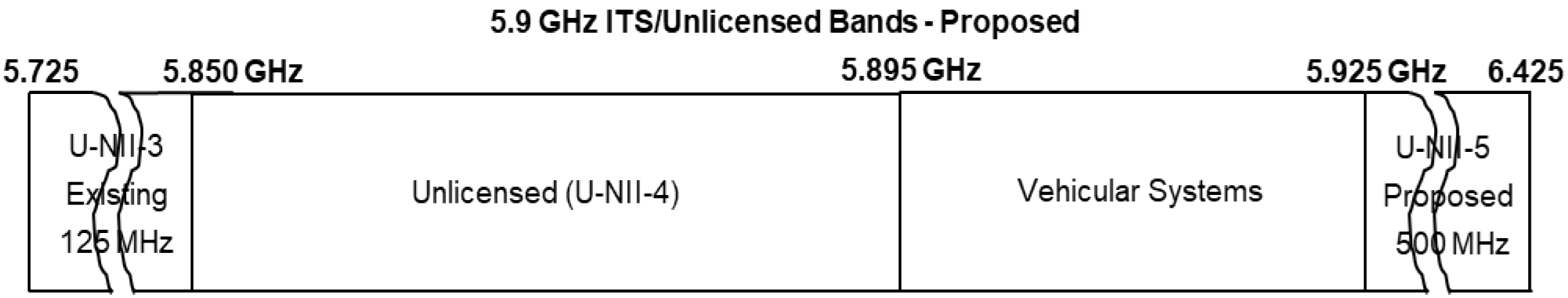}
\caption{FCC's 5.9 GHz band reform \cite{rns23_2}}
\label{fig_0}
\end{figure}

\section{Related Work}\label{sec_related}
\subsection{Performance Analysis Schemes}
\subsubsection{Mathematical Analysis Framework}
Analysis frameworks based on stochastic geometry for DSRC have been provided recently \cite{arxiv19}-\cite{access19}. They commonly rely on the fact that uniform distributions of nodes on $X$ and $Y$ axes of a Cartesian-coordinate two-dimensional space yield a Poisson point process (PPP) on the number of nodes in the space \cite{haenggi05}. This paper also applies the stochastic geometry framework for analysis of the proposed mechanism.

\subsubsection{Performance Evaluation Method}
A recent proposal combines a packet-level simulation model with data collected from an actual vehicular network \cite{gdot_13}; however, the potential impacts of internal and external bandwidth contention were not studied, which forms a critical discussion point after the US FCC's recent 5.9 GHz band reform \cite{rns23_2}. For instance, it is assumed that safety messages and Internet packets are sent over separate DSRC channels \cite{gdot_13}, whereby no interference is generated between safety and Internet traffic. This assumption has become obsolete according to the US FCC's recent proposition where DSRC is unable to utilize multiple channels any more \cite{rns23_2}.

\subsubsection{DSRC in High Traffic Density}
It has been found that a DSRC network is more constrained by packet expirations rather than collisions over the air. Thus, we put particular focus on the performance of a DSRC network in a high density of traffic. The performance of a DSRC broadcast system in a high-density vehicle environment has been studied \cite{gdot_11}, yet the assumption was too ideal to be realistic--i.e., the number of vehicles within a vehicle’s communication range was kept constant. Another study proposed a DSRC-based traffic light control system \cite{gdot_12}, but it limited the applicability to the traffic lights only.

\subsubsection{Safety-Related Application}
Furthermore, we concentrate on DSRC's networking to support the safety-critical applications. In the related literature, a DSRC-based end of queue collision warning system has been proposed \cite{gdot_14}. However, it discusses a one-dimensional freeway model, which needs significant improvement for application to an intersection with two or more ways.

\subsubsection{External Bandwidth Contention}
Lastly, the objective of our proposed protocol is to lighten the traffic load of a DSRC network to better suit in an environment of coexisting with a disparate technology (i.e., C-V2X) according to the 5.9 GHz reform \cite{rns23_2}.  The performance degradation of DSRC under interference from Wi-Fi has been studied \cite{gdot_15}; however, it lacks consideration of coexistence with C-V2X.

\subsection{Performance Improvement Schemes}
In general IEEE 802.11 carrier-sense multiple access (CSMA), various modifications on the binary exponential backoff (BEB) algorithm have been tried as a means to improve throughput and fairness. Specifically, adjustment of the CW was often suggested to improve the performance of a vehicular communications network such as a recent work \cite{wu2018improving}. More directly relevant to our work, a distance based routing protocol has been found to perform better in vehicular ad-hoc networks (VANETs) \cite{ramakrishna2012dbr}. Also, in a general ad-hoc network, reduction of the length of a header can be a solution that is worth considering \cite{grasnet}; however, due to being a centralized architecture, it shows a limit to be applied to a V2X network. A ``subjective'' user-end experience optimization is also worth consideration \cite{gilsoo}, wherein a one-bit user satisfaction indicator was introduced, which served as the objective function in a non-convex optimization.

\begin{figure}
\begin{minipage}{\linewidth}
\centering
\includegraphics[width = \linewidth]{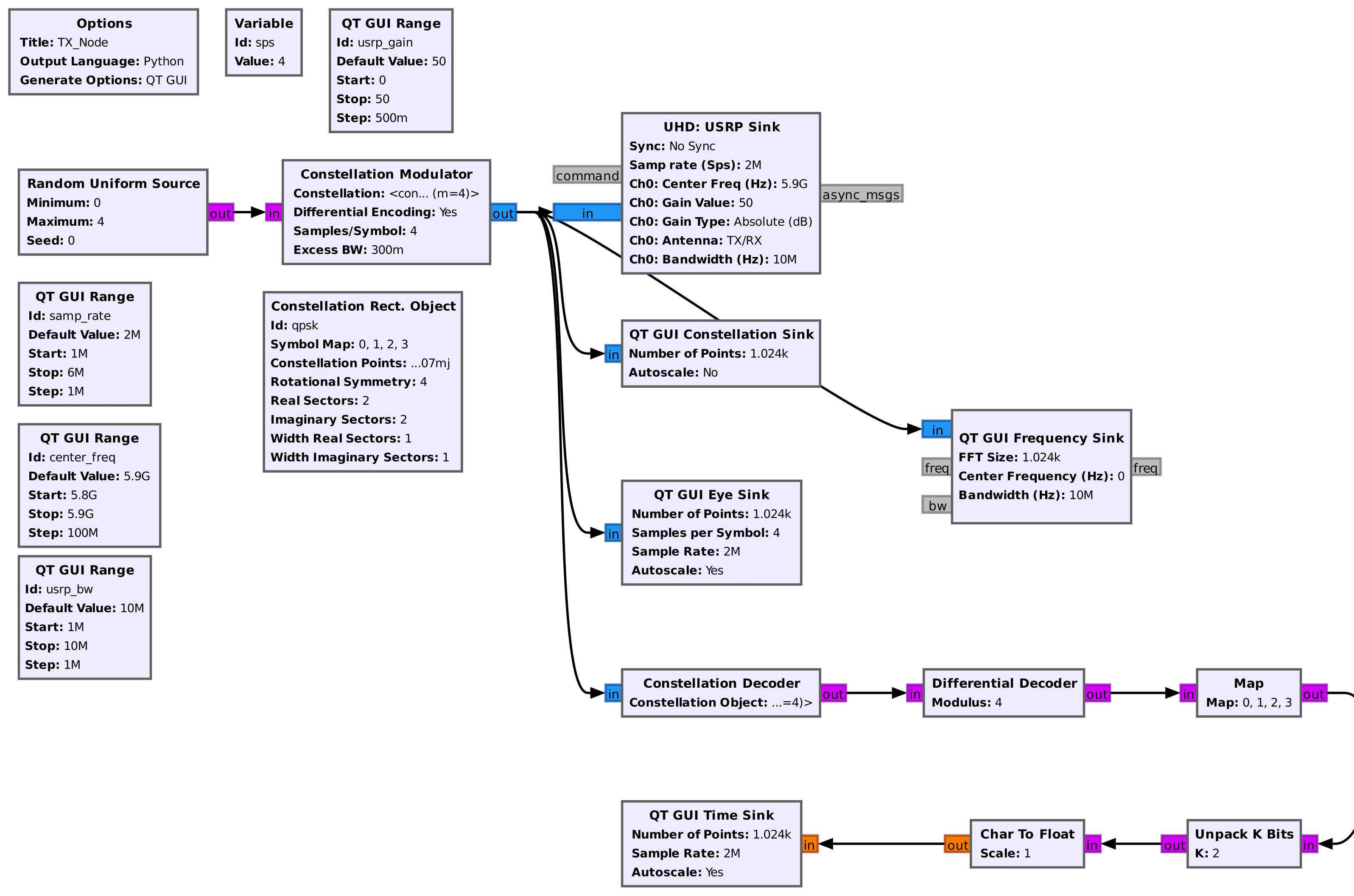}
\caption{Flowgraph for TX node on GNU Radio Companion}
\label{fig_1}
\end{minipage}
\begin{minipage}{\linewidth}
\vspace{0.3 in}
\centering
\includegraphics[width = \linewidth]{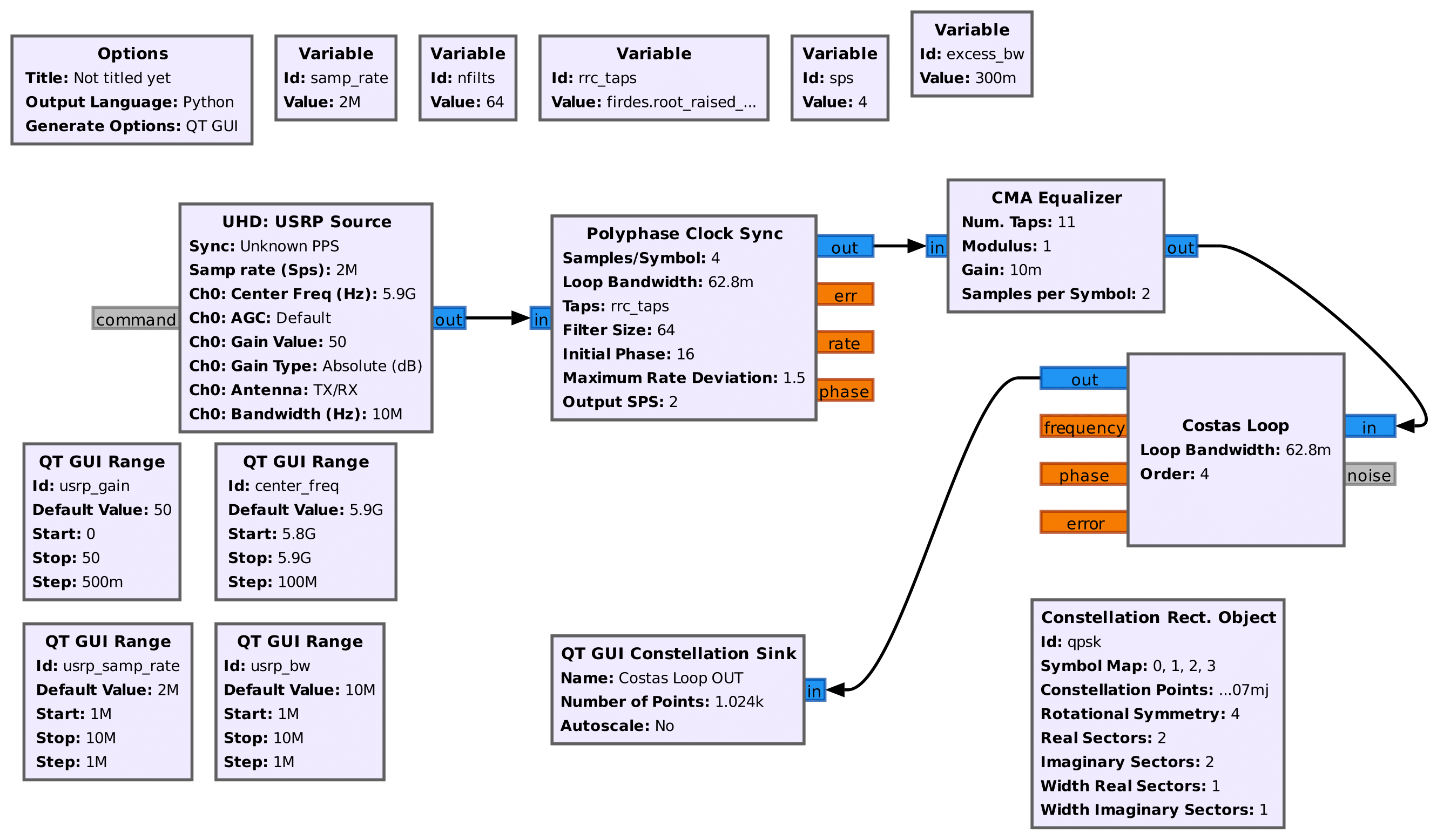}
\caption{Flowgraph for RX node on GNU Radio Companion}
\label{fig_2}
\end{minipage}
\end{figure}

\section{System Model}\label{sec_model}
A primitive radio communication link was established by means of two software defined radios (SDR) in which one radio was configured to exclusively make transmissions while the second radio exclusively received and recovered the transmitted signals. The SDR's used to implement the link were the National Instrument (NI) USRP-2954R and the Ettus USRP X310 (each equipped with a UBX-160 daughterboard) of which were controlled and interfaced with by means of the software known as GNU Radio Companion installed on Raspberry Pi 4B (8 GB RAM) microprocessors running Raspbian operating system (OS), a derivative of Debian Linux distribution. A single VERT 2450, dual-band, vertical antenna was attached to each radio to facilitate transmissions at a center frequency of 5.9 GHz, and the MATLAB computing platform was utilized as the primary means for analyzing/post-processing data stored by the receiver (RX) node during testing. 

GNU Radio Companion was selected for interacting with the SDR's due to its ease of use and the abundant amount of information already available for reference on the official GNU Radio Wiki and other engineering forums alike. Informational guides such as ``QPSK Mod and Demod'' provided on the wiki page helped to form the foundation upon which the investigations described by this report were performed \cite{gnu_qpsk}. The GNU Radio flowgraphs utilized during testing of the transmitter (TX)-RX link were derived directly from the aforementioned guide and have been adapted to accommodate the SDR hardware as required. 

Testing was performed by connecting each of the SDR's to an assigned Raspberry Pi via ethernet and executing python programs produced subsequent to the implementation of the TX and RX flowgraphs in GNU Radio Companion (see Figs. \ref{fig_1} and \ref{fig_2} below). The node responsible for transmitting performed QPSK modulation on an input before providing the modulated data to a ``USRP Sink'' block, at which point the signal was transmitted by the SDR at a carrier frequency of 5.9 GHz with a bandwidth of 10 MHz. Two different data sources were utilized while investigating the link, the ``Random Uniform Source'' block, and the ``Vector Source'' block. Initial testing exclusively relied upon the Random Uniform Source of which generated random samples output in bytes with values between the parameters provided to the block. Similarly, the Vector Source was used to produce a stream of data output in bytes, however, unlike the Random Uniform Source, the data provided by the Vector Source was generated in the same order in which it was input to the block.

The flowgraph implemented to facilitate the recovery of transmitted signals began with a ``USRP Source'' block of which allowed for GNU Radio Companion to interact with the SDR being utilized as the receiver in the link. The USRP Source block was tuned to the same carrier frequency as that of the transmitter, 5.9 GHz, and also featured a bandwidth of 10 MHz. Signals received by the SDR were subjected to a recovery loop within the RX flowgraph prior to reaching the demodulation block/outputs (spectrum analyzer, constellation diagrams, file sink, etc..). The recovery process was absolutely critical as transmitted signals are known to experience distortion deriving from sources such as inter-symbol interference and multipath losses while traveling through a channel to the receiver. Signal recovery was facilitated by a series of blocks meant to correct distortions in the timing, phase, and frequency of the received signal such as the Polyphase Clock Sync, the constant modulus algorithm (CMA) Equalizer, and the Costas Loop.

All testing was performed with the SDR's positioned approximately two feet apart from antenna to antenna as to mitigate losses to the maximum extent possible without needing to exceed the gain limits of the radios defined by the aforementioned USRP Source and Sink blocks. The environment in which the tests were performed included standard appliances (tv's, monitors, Wi-Fi Router, etc..) within a close proximity, however, no nearby devices were known to be operating at the 5.9 GHz range during experimentation. Testing was always initiated by executing the transmitting program implemented at the TX node. Once transmissions had begun, the program that facilitated the recovery of the transmitted signals was executed. For tests that involved storing data, the recovery program was terminated after approximately 10 seconds worth of run time due to the high volume of data that was typically stored during each trial. The results of the tests were captured/stored for post-processing and analysis.

\begin{figure}
\begin{minipage}{\linewidth}
\centering
\includegraphics[width = \linewidth]{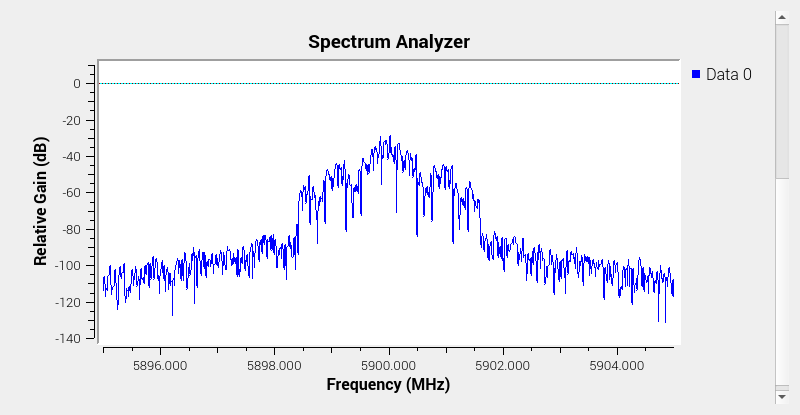}
\caption{Random Uniform Source: Spectrum analyzer output}
\label{fig_3}
\end{minipage}
\begin{minipage}{\linewidth}
\vspace{0.3 in}
\centering
\includegraphics[width = \linewidth]{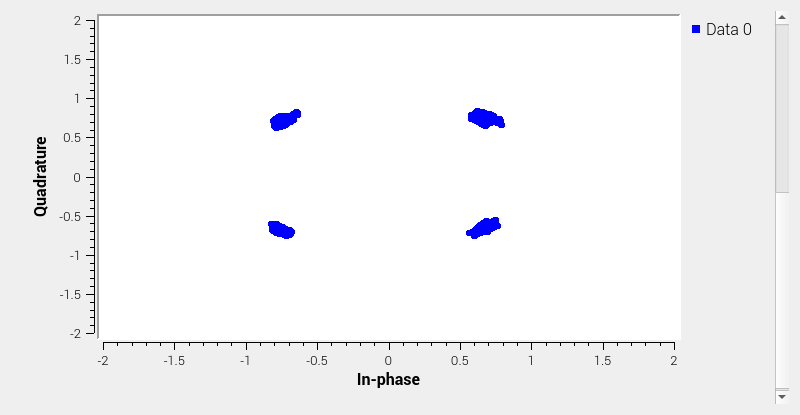}
\caption{Random Uniform Source: Constellation diagram}
\label{fig_4}
\end{minipage}
\begin{minipage}{\linewidth}
\vspace{0.3 in}
\centering
\includegraphics[width = \linewidth]{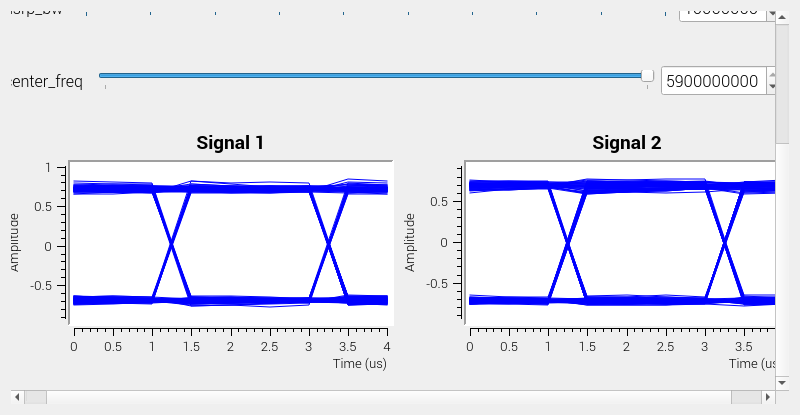}
\caption{Random Uniform Source: Eye diagram}
\label{fig_5}
\end{minipage}
\end{figure}

\section{Experimental Results}\label{sec_results}
The experimental results produced subsequent to testing the radio link are given by spectrum analyzer outputs, eye diagrams, constellation diagrams, and raw binary data of which have been categorized with respect to the data source utilized during the trial from which the given result derives from. As such, the results of the investigation thus far were produced when either the Random Uniform Source or the Vector Source blocks provided input to the TX node. Given below are the plots produced by the receiver and the binary data (Vector Source case only) that was stored and evaluated using the MATLAB software environment. 

Figs. \ref{fig_3}-\ref{fig_5} display the spectrum, constellation diagram, and eye diagram produced by the receiver during recovery of a transmitted signal when the input to the TX node was generated by the \textit{Random Uniform Source} block with a range of values: [0,4). Each of the diagrams represent the transmitted signal as expected with the constellation diagram (Fig. \ref{fig_4}) clearly displaying four distinctly mapped symbols of which is a prominent characteristic of a QPSK modulation scheme.

\begin{figure}
\begin{minipage}{\linewidth}
\centering
\includegraphics[width = \linewidth]{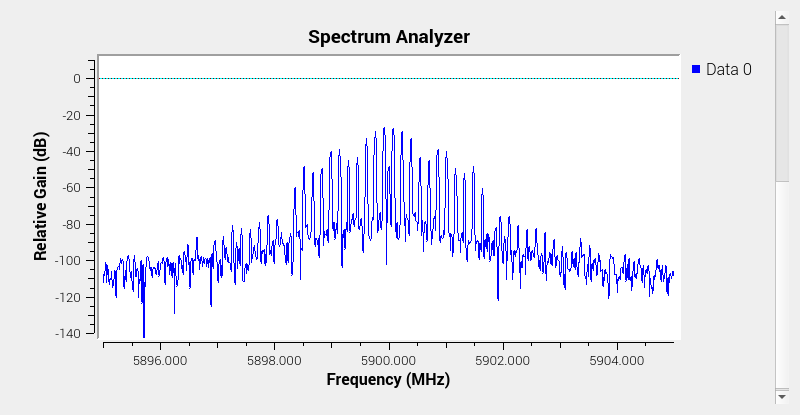}
\caption{Vector Source--Sequence 1: Spectrum analyzer output}
\label{fig_6}
\end{minipage}
\begin{minipage}{\linewidth}
\vspace{0.3 in}
\centering
\includegraphics[width = \linewidth]{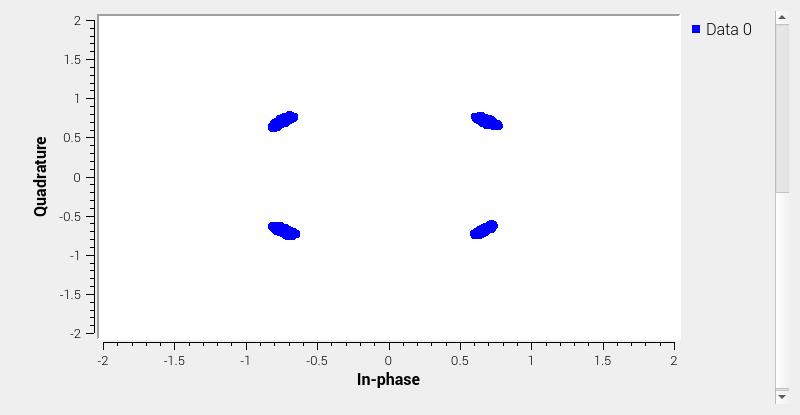}
\caption{Vector Source--Sequence 1: Constellation diagram}
\label{fig_7}
\end{minipage}
\begin{minipage}{\linewidth}
\vspace{0.3 in}
\centering
\includegraphics[width = \linewidth]{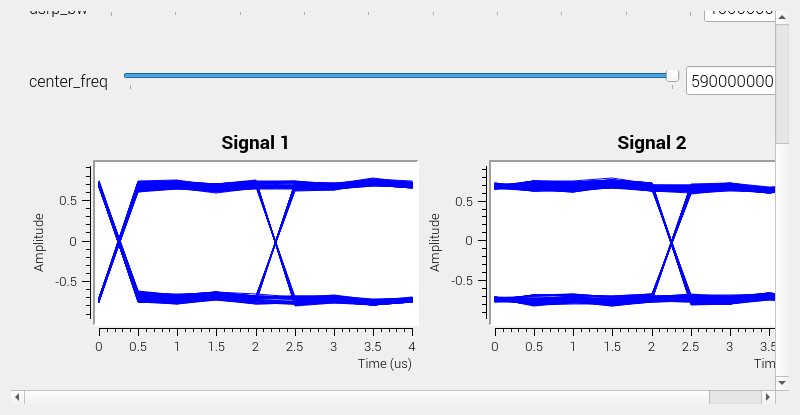}
\caption{Vector Source--Sequence 1: Eye diagram}
\label{fig_8}
\end{minipage}
\end{figure}

Similarly, Figs. \ref{fig_6}-\ref{fig_8} display the results produced when the \textit{Vector Source} block was used to provide input sequence, 0, 1, 2, 3 (Sequence 1), to the TX node. As expected, the spectrum and constellation/eye-diagrams resembled that of those produced during the trials involving the Random Uniform Source block, each of which are indicative of a successful transmission/recovery.

\begin{figure}
\centering
\includegraphics[width = \linewidth]{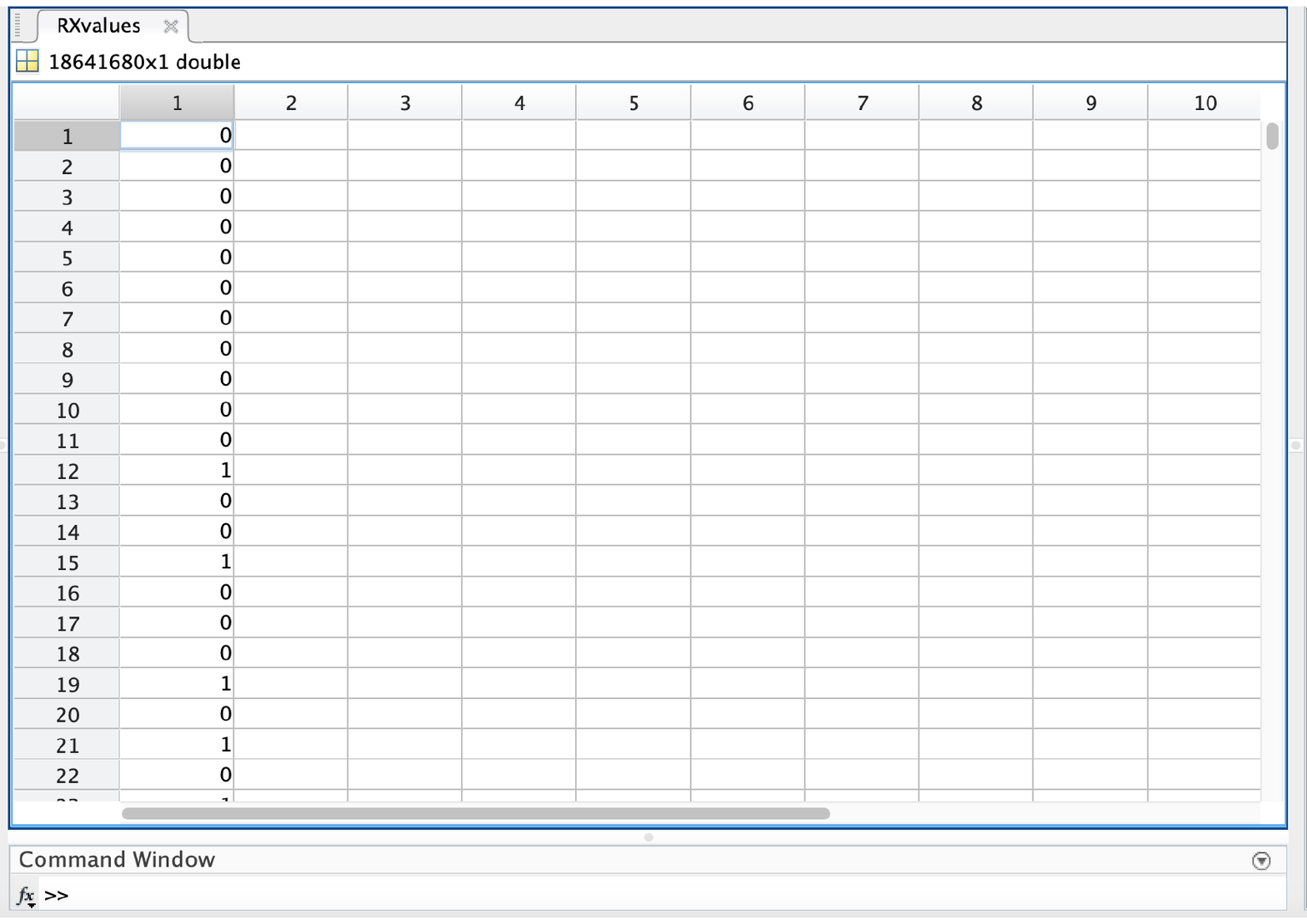}
\caption{Vector Source--Sequence 1: Binary data pre-signal synchronization}
\label{fig_9}
\end{figure}

\begin{figure}
\centering
\includegraphics[width = \linewidth]{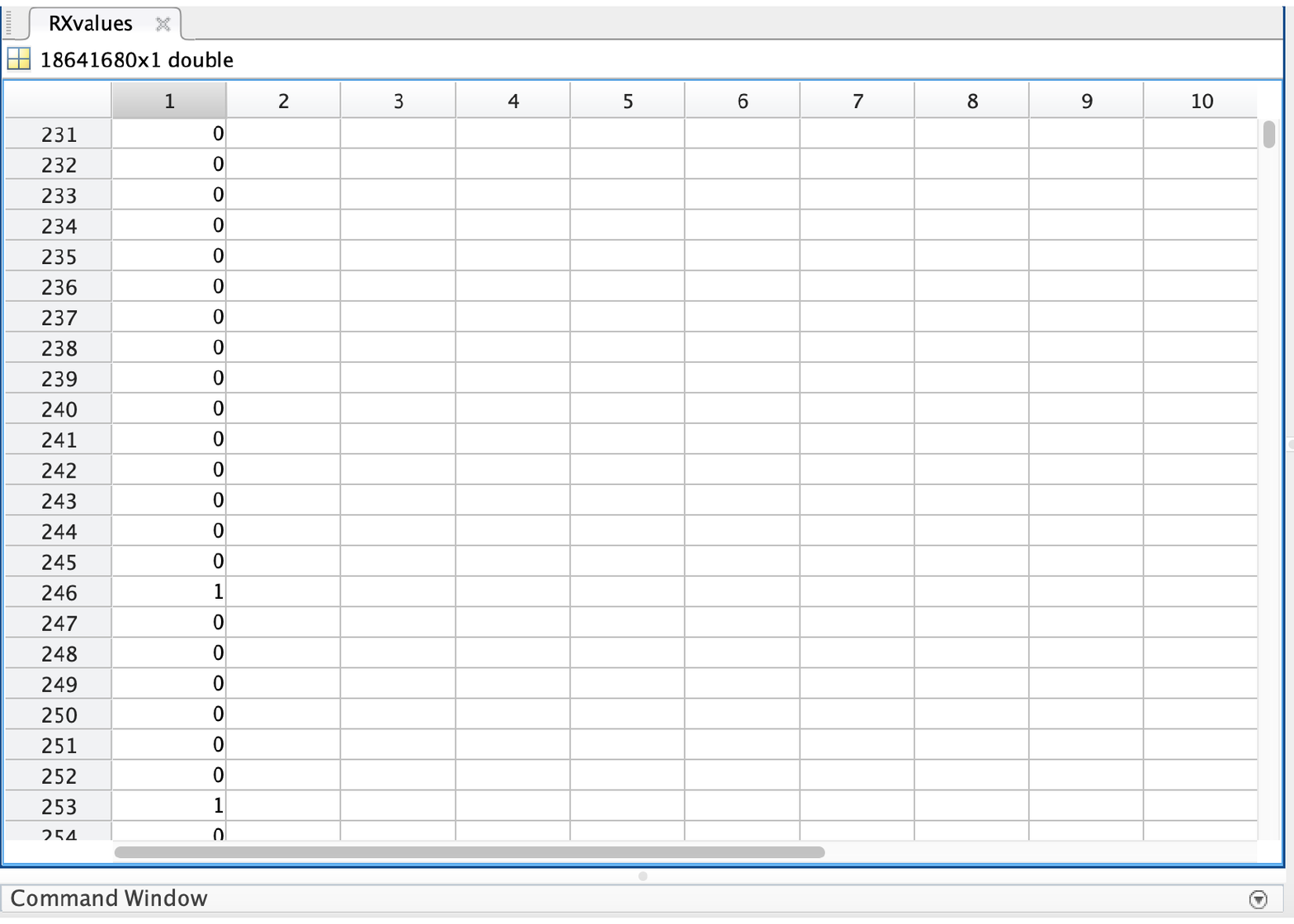}
\caption{Vector Source--Sequence 1: Binary data synchronized signal (Part 1)}
\label{fig_10}
\end{figure}

\begin{figure}
\centering
\includegraphics[width = \linewidth]{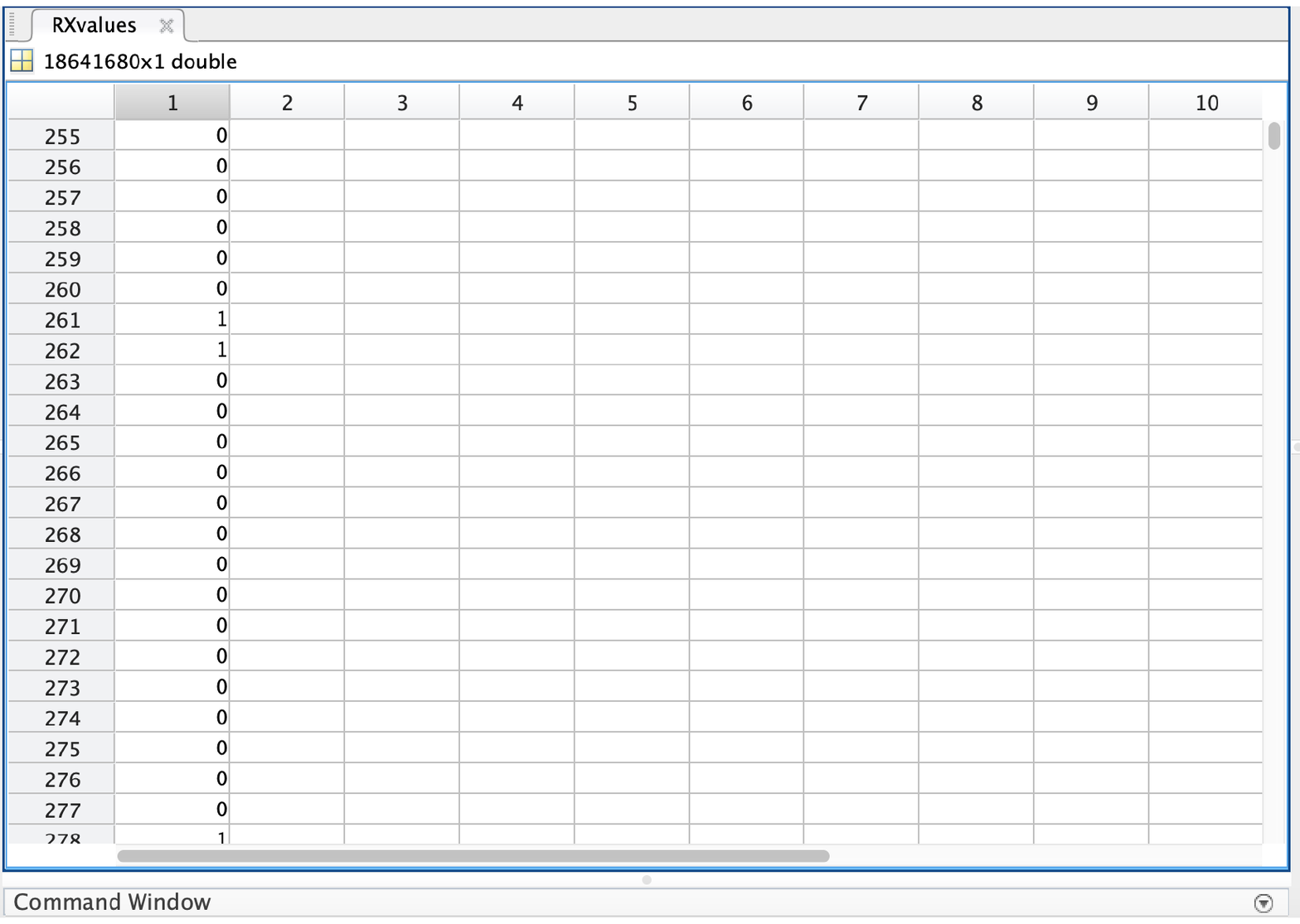}
\caption{Vector Source--Sequence 1: Binary data synchronized signal (Part 2)}
\label{fig_11}
\end{figure}

To further verify that the data being transmitted by the link was being recovered successfully, a sample of the recovered data was stored during one of the trials of the experiment involving Vector Source Sequence 1 (0, 1, 2, 3). Subsequent to accessing the binary file produced by the receiver using MATLAB, it was confirmed that the sequence had been transmitted and recovered as anticipated. Figs. \ref{fig_9}-\ref{fig_11} display the binary data stored by the receiver during such trials with Figs. \ref{fig_10} and \ref{fig_11} displaying one full iteration of Sequence 1. The image given by Figure 9 reveals that the sequence was not correctly recovered when the receiver was first activated. This is a significant observation as it provides important insight into how a QPSK communication system sometimes operates. In the case of the recovery loop implemented at the RX node (see Fig. \ref{fig_2}), a short period of time was required to pass while the loop synchronized with the transmitted signal before the recovered data could be recognized as the original sequence.

In attempt to expand upon the capabilities of the still very primitive communication system, another sequence (denoted as Sequence 2) was passed to the Vector Source block as follows: 0, 255, 72, 101, 108, 108, 111, 87, 111, 114, 108, 100. The first two values in the sequence, 0 and 255, were included as an attempt to mark the beginning of each iteration of which could be identified during post-processing analysis in MATLAB. The rest of the sequence can be translated using a standard ASCII table to reveal the message, ``Hello World.'' The results of the tests involving Sequence 2 are given by Fig. \ref{fig_12}-\ref{fig_14} in which it is clear that the signal was not recovered correctly. The constellation and eye diagrams reveal a considerable amount of distortion, and the binary data (not pictured) did not exhibit any recognizable patterns in line with the predetermined sequence.

\begin{figure}
\begin{minipage}{\linewidth}
\centering
\includegraphics[width = \linewidth]{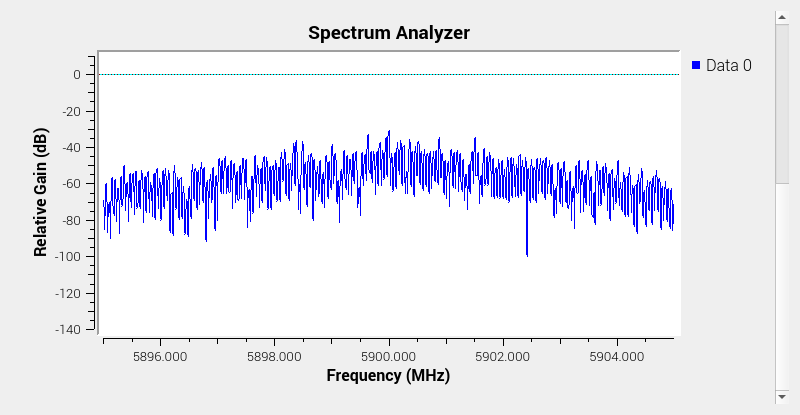}
\caption{Vector Source--Sequence 2: Spectrum analyzer output}
\label{fig_12}
\end{minipage}
\begin{minipage}{\linewidth}
\vspace{0.3 in}
\centering
\includegraphics[width = \linewidth]{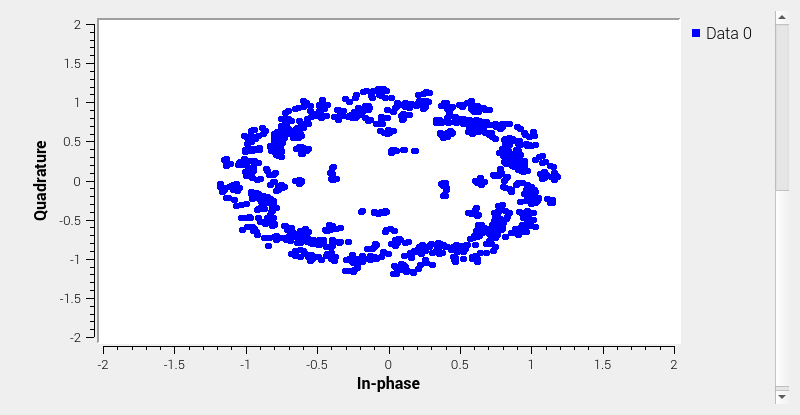}
\caption{Vector Source--Sequence 2: Constellation diagram}
\label{fig_13}
\end{minipage}
\begin{minipage}{\linewidth}
\vspace{0.3 in}
\centering
\includegraphics[width = \linewidth]{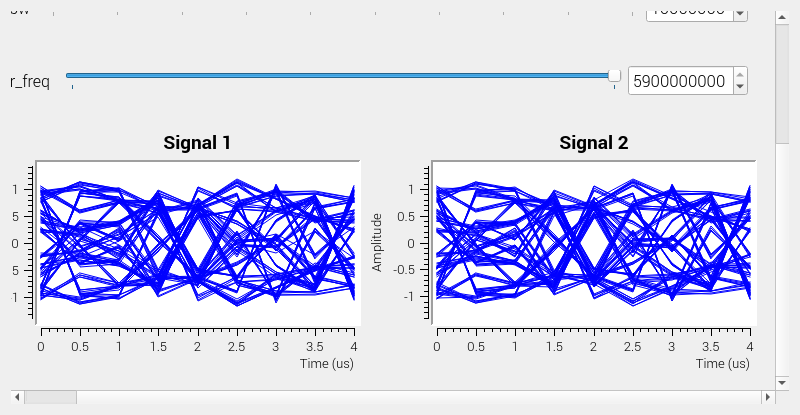}
\caption{Vector Source--Sequence 2: Eye diagram}
\label{fig_14}
\end{minipage}
\end{figure}

\section{Concluding Remarks}\label{sec_conclusions}
The findings of the tests performed using two USRP software defined radios confirm that it is possible to implement a basic communications link with the help of the GNU Radio Companion software. It was successfully shown that small sequences of binary data could be transmitted and recovered using a QPSK modulation scheme despite the current inability of the link to transmit more meaningful information.

Future investigations will attempt to address the reasons for the distortions observed in the transmissions recovered depicted by this paper's results with the goal of the project being the development of a functional communication system. Furthermore, the feasibility of advanced techniques as a method to improve the performance will be investigated such as reinforcement learning \cite{mine_tiv23}, blockchain \cite{mine_iceic22}.



\end{document}